\documentclass{iau} 
\usepackage{graphicx}

\title[H$_2$ content of galaxies inside and around intermediate redshift clusters] 
{H$_2$ content of galaxies inside and around intermediate redshift clusters}

\author[Damien Sp\'{e}rone-Longin]   
{Damien Sp\'{e}rone-Longin}

\affiliation{Laboratoire d'Astrophysique, Ecole Polytechnique F\'{e}d\'{e}rale de Lausanne (EPFL), \\ 1290 Sauverny, Switzerland \\ email: {\tt damien.sperone-longin@epfl.ch}}

\pubyear{2020}
\volume{359}  
\jname{Galaxy evolution and feedback across different environments}
\editors{T. Storchi-Bergmann, R. Overzier, W. Forman \& R. Riffel, eds.}

\begin{document}

\maketitle

\begin{abstract}
Dense environments have an impact on the star formation rate of galaxies. 
As stars form from molecular gas, looking at the cold molecular gas content 
of a galaxy gives useful insights on its efficiency in forming stars. 
However, most galaxies observed in CO (a proxy for the cold molecular 
gas content) at intermediate redshifts, are field galaxies. Only a handful of 
studies focused on cluster galaxies.
I present new results on the environment of one medium mass 
cluster from the EDisCS survey at $z\sim0.5$. 27 star-forming galaxies were 
selected to evenly sample the range of densities encountered inside and 
around the cluster. We cover a region extending as far as 8 virial radii from 
the cluster center. Indeed there is ample evidence that star formation 
quenching starts already beyond 3 cluster virial radii. I discuss our 
CO(3-2) ALMA observations, which unveil a large fraction of galaxies with 
low gas-to-stellar mass ratios.
\keywords{galaxies: evolution -- galaxies: clusters: general -- submillimeter: galaxies}
\end{abstract}

\firstsection 

\section{SEEDisCS}

We initiated the first systematic study of galaxy properties along the large scale structures
(LSS) feeding galaxy clusters, so called Spatially Extended EDisCS (SEEDisCS).
It focuses on 2 clusters from the ESO Distant Cluster Survey (EDisCS, 
\cite[White et al. 2005]{White2005}):
CL1301.7$-$1139 and CL1411.1$-$1148. They are located at  redshifts $z_{\rm cl} = 0.4828$
and 0.5195 and have velocity dispersions of $\sigma_{\rm cl} = 681$ and 710 km/s,
respectively. Deep $u, g, r, i, z$ and $K_{\rm s}$ images were gathered with 
CFHT/MEGACAM and WIRCam. They cover a region that extends up to around 10 times the cluster virial radius, $10\,R_{200}$.
This survey is divided into 3 main steps. The first one consists of the identification
of the large scale structures around the clusters thanks to accurate photometric
redshifts. The second is a spectroscopic follow-up of these large scale structures to study
the properties of the galaxy stellar populations. The last step is made up of 
ALMA programs to study the cold molecular gas reservoir status of galaxies.
Here, we focus on one cluster : CL1411.1$-$1148, since analysis is still underway for the other one.

\section{ALMA Sample}

\begin{figure}
\centering
\includegraphics[width=0.75\textwidth]{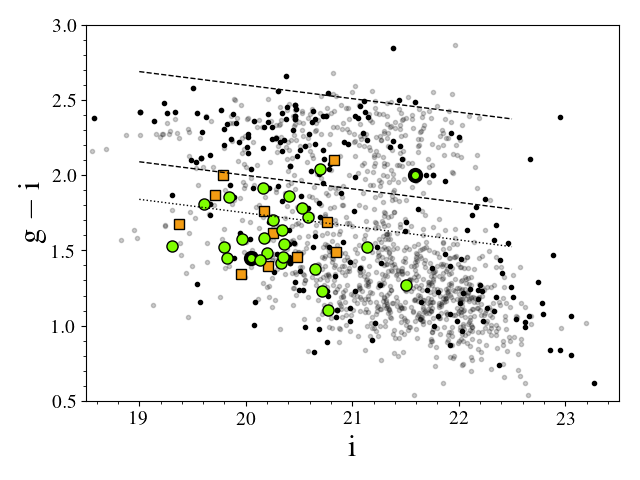}
\caption{Observed colour-magnitude diagram, $g-i$ as a function of $i$ for 
the CL1411.1$-$1148 galaxies. Our ALMA sample is represented by the circles. 
The markers with the larger edges are for our \textit{Spitzer}-observed galaxies. 
The squares are for parts of the PHIBSS2 sample. 
The small gray dots are galaxies with photometric redshifts and the black ones 
are for galaxies with spectroscopic redshifts, within the cluster limit. 
The dashed lines delimit the red sequence ($\pm$0.3 mag above 
\cite[De Lucia et al. (2007)]{DeLucia2007} relation) and the dotted line 
delimits the transition zone between the blue clump and the red sequence 
(0.3 mag below the red sequence).}
\label{CMD}
\end{figure}  

We selected 27 star-forming galaxies from the spectroscopic sample
within $5\times R_{200}$ of CL1411.1$-$1148 to be observed with ALMA. 
This is the largest sample of galaxies with direct cold gas measurements 
at intermediate redshift and the only one of galaxies in interconnected 
cosmic structures around a galaxy cluster.
Those galaxies were selected to have similar colours as the 
PHIBSS2 \textit{field} main-sequence star-forming galaxies sample 
(\cite[Freundlich et al. 2019]{Freundlich2019}) with redshift between $z=0.5$ and 0.6.
Figure \ref{CMD} shows the position of our targets in the $g - i$ vs $i$ diagram 
with respect to the other cluster galaxies and the PHIBSS2/COSMOS sample. 
Those 27 targets also sample the range of densities encountered inside 
and around the cluster.
They were observed at 226 GHz to detect the CO${_{J=3\rightarrow2}}$ transition. 
We benefited from 23h of observations during Cycle 3 and 5, and reported 27 detections.

\begin{figure}
\centering
\includegraphics[width=0.75\textwidth]{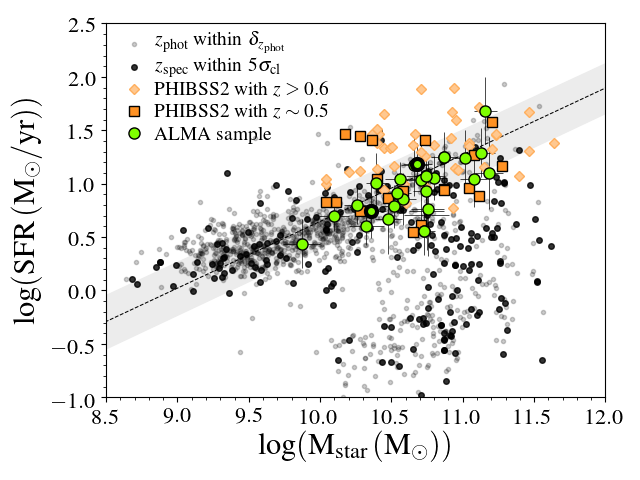}
\caption{Location of the CL1411.1$-$1148 (dots) and ALMA (circles) galaxies 
in the stellar mass-SFR plane. The markers with thicker black edges are for our
\textit{Spitzer}-observed galaxies.
The squares are the PHIBSS2 galaxies at our cluster's redshift and the 
diamonds are the rest of the low-$z$ PHIBSS2 sample ($0.6 < z \leq 0.8$). 
The middle line is the \cite[Speagle et al. (2014)]{Speagle2014} 
main-sequence at our cluster redshift with the corresponding $\pm$0.3dex 
scatter as the shaded area.}
\label{MS}
\end{figure}

The star formation rates (SFR) and stellar masses (M$_{\rm star}$) were derived 
using \texttt{MAGPHYS} (\cite[da Cunha et al. 2008]{daCunha2008}) 
for our targets and all the galaxies within the cluster redshift limits. 
Figure \ref{MS} shows the SFR-M$_{\rm star}$ diagram with the position of 
our ALMA sample, circles, compared to the position of the PHIBSS2 sample, squares and diamonds.
The limits in M$_{\rm star}$ and SFR are similar for both samples, as expected from our colour-based selection. 
Hence, our galaxies are located on the main-sequence or slightly below. 
Our sample is also evenly sampling the main-sequence between $10.3 \leq \log({\rm M_{star} / M_\odot}) \leq 11.3$.

From the ALMA observations, the CO luminosities were derived following the
\cite[Solomon \& Vanden Bout (2005)]{Solomon2005} recipe.
To derive the cold molecular gas masses, M$_{\rm H_2}$, we used 
\begin{equation} \label{eq1}
    M_\mathrm{H_2}=\alpha_\mathrm{CO} \frac{L'_\mathrm{CO(3\rightarrow 2)}}{r_{31}}, 
\end{equation}
where $r_{31} = 0.5$ is the conversion factor from the third rotational transition of 
CO to the first (\cite[Genzel et al. 2015]{Genzel2015}; \cite[Chapman et al. 2015]{Chapman2015};
\cite[Carleton et al. 2017]{Carleton2017} and \cite[Tacconi et al. 2018]{Tacconi2018}), and
$\alpha_{\rm CO} = \alpha_{\rm MW} = 4.36\,{\rm M_\odot(K\,km/s\,pc^2)^{-1}}$
is the CO(1-0) luminosity-to-molecular-gas-mass conversion factor, considering a 36\% correction to account for interstellar helium (\cite[Leroy et al. 2011]{Leroy2011}; \cite[Bolatto et al. 2013]{Bolatto2013}; 
\cite[Carleton et al. 2017]{Carleton2017}).

\section{Results}
\begin{figure}
\centering
\includegraphics[width=0.75\textwidth]{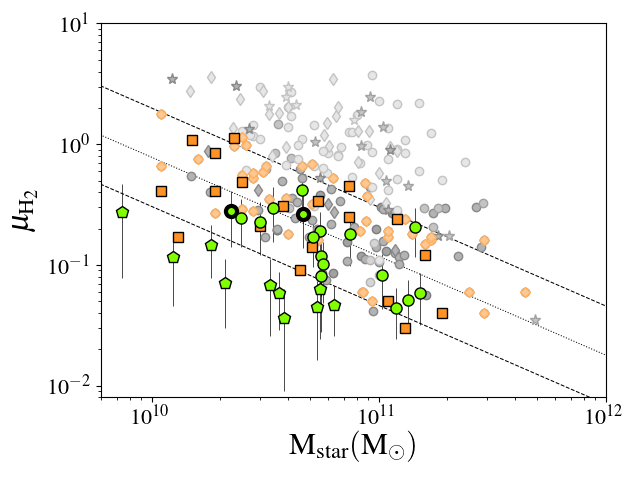}
\caption{Fraction of cold molecular gas with respect to stellar masses. The circles 
and pentagons are our ALMA sample. The squares and diamonds are the same
as in Fig. \ref{MS}. The gray markers in the background are other galaxies from
the literature. The dotted line is the fit of the M$_{\rm star}$-$\mu_{\rm H_2}$ relation.
The dashed line is its dispersion. The pentagons are our galaxies with cold gas
ratios below the dispersion of the previous relation.}
\label{muh2_mstar}
\end{figure}  

In order to compare our results to the literature and the PHIBSS2 sample, we used 
the cold molecular gas-to-stellar mass ratio: $\mu_{\rm H_2}= {\rm M_{H_2}/M_{star}}$.
Figure \ref{muh2_mstar} shows the cold molecular gas-to-stellar mass ratio as 
a function of the stellar mass for our galaxies and comparison samples.
One can see that while 67\% of our targeted galaxies have comparable gas fractions 
as their field counterparts at the same redshift, 33\% are populating a new area of 
low $\mu_{\rm H_2}$.
They are represented by pentagons in Fig. \ref{muh2_mstar}. They fall below 
the $1\sigma$ dispersion of the relation between $\mu_{\rm H_2}$ and M$_{\rm star}$,
which was derived using the PHIBSS2 sample at $z\sim 0.5$.
Their large number indicates they are not simply part of the tail of the distribution of
the field galaxies but rather a new, different population, which implies a different relation
between $\mu_{\rm H_2}$ and M$_{\rm star}$, specifically in the $10.2 \leq 
\log({\rm M_{star} / M_\odot}) \leq 10.85$
mass range.

\begin{figure}
\centering
\includegraphics[width=0.75\textwidth]{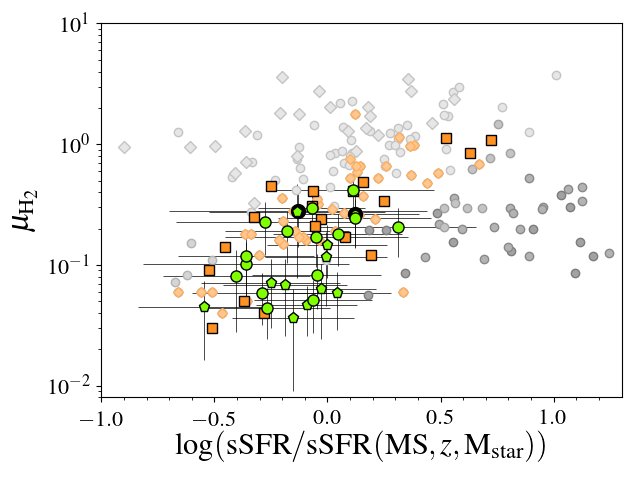}
\caption{Fraction of cold molecular gas with respect to specific SFRs. Markers are the 
same as on Fig. \ref{muh2_mstar}}
\label{muh2_ssfr}
\end{figure}  

Figure \ref{muh2_ssfr} shows how the cold gas ratio scales with the specific SFR (sSFR), 
normalized to the one of the main-sequence, of galaxies at different redshifts. 
At $z\sim0.5$, a trend can be noticed. Most of our targets follow it, but here again, 
some of them are populating a new area, which corresponds to galaxies 
lying within or very close to the main-sequence but depleted in cold molecular gas.

A question we can ask is whether the values chosen for the different factors used
to derive M$_{\rm H_2}$ can have an influence on our $\mu_{\rm H_2}$.

\underline{$\alpha_{\rm CO}$} : we chose to use the same coefficient as for the 
Milky-Way, whereas, in PHIBSS2, \cite[Freundlich et al. (2019)]{Freundlich2019} 
are using a coefficient based on the mass-metallicity relation derived by 
\cite[Genzel et al. (2015)]{Genzel2015}. If we were to use this recipe,
we would have $3.8 \leq \alpha_{\rm CO} \leq 4.9$, which would lower
our already low $\mu_{\rm H_2}$.

\underline{$r_{31}$} : We took the same value as Genzel et al. (2015). But
\cite[Dumke et al. (2001)]{Dumke2001}, by looking at local spirals, showed that 
$r_{31}$ could vary within a spiral galaxy. Indeed, it would be closer to 0.8 in 
the center of those galaxies, which would lower our ratios by 38\%. 
This coefficient can be closer to 0.4 in the outer parts of the disk of spiral galaxies, 
which would lead to increase our ratios by 25\%, but this is not enough 
to put our low $\mu_{\rm H_2}$ back into the "normal" sequence.

Moreover, those low gas fraction galaxies have narrower CO line widths compared to 
their PHIBSS2 counterparts. Therefore, the former potentially have more extended 
morphologies ($\Delta V\propto$ Mass/Radius) than the normal $\mu_{\rm H_2}$ and field galaxies.
This could be a signature of environmental effects at play in the interconnected structures surrounding the cluster.

\section{Summary}
We report the first large sample of galaxies with direct cold molecular gas measurements 
within the same cluster environment at intermediate redshift. 
Those 27 galaxies were selected to have the same colours as the PHIBSS2 galaxies. 
No prior selection on SFR was applied as it is usually the case for such surveys and larger.
A large portion, 67\%, of our targets have similar gas content as other field galaxies (PHIBSS2) at
$z\sim0.5$. The rest of them, 33\%, have low cold gas to stellar mass ratio despite being main-sequence
star-forming galaxies. The latter have lower CO line widths than their field counterparts, which is an
indication of more extended morphologies.
To summarize, we unveiled a new population of low gas content star-forming galaxies, using a
different selection criterion, and located in the different environments encountered around a cluster.

\end{document}